\documentclass{article}
\usepackage{algpseudocode}
\usepackage{algorithm}
\usepackage{bm}
    \PassOptionsToPackage{numbers, compress}{natbib}
\usepackage[symbol]{footmisc}

\usepackage[preprint]{neurips_2024}
\usepackage{wrapfig}

\usepackage[utf8]{inputenc} 
\usepackage[T1]{fontenc}    
\usepackage{hyperref}       
\usepackage{url}            
\usepackage{booktabs}       
\usepackage{amsfonts}       
\usepackage{nicefrac}       
\usepackage{microtype}      
\usepackage{xcolor}         
\usepackage{mathtools, amsmath, amssymb, amsthm}
\usepackage{makecell}
\usepackage{graphicx} 
\usepackage{array}    
\usepackage{caption}  
\usepackage{wrapfig}
\usepackage{sidecap}
\newcommand{\chroma}{\texttt{Chroma}}

\DeclareRobustCommand{\rchi}{{\mathpalette\irchi\relax}}
\newcommand{\irchi}[2]{\raisebox{\depth}{$#1\chi$}} 

\title{Generative modeling of protein ensembles guided by crystallographic electron densities}

\author{
{Sai Advaith Maddipatla\thanks{Equal contribution. Correspondence Email: \texttt{\{maddipatlas, n.sellam\}@campus.technion.ac.il}}}\\Technion, Israel \\
\And {Nadav Bojan Sellam$^\ast$}\\Technion, Israel \\
\And {Sanketh Vedula}\\Technion, Israel \\
\And {Ailie Marx}\\Migal -- Galilee Research Institute, Israel \\
\And {Alex Bronstein}\\Technion, Israel and IST Austria\\
}

\begin{document}

\maketitle

\begin{abstract}
Proteins are dynamic, adopting ensembles of conformations.  The nature of this conformational heterogenity is imprinted in the raw electron density measurements obtained from X-ray crystallography experiments.  Fitting an ensemble of protein structures to these measurements is a challenging, ill-posed inverse problem. We propose a non-i.i.d. ensemble guidance approach to solve this problem using existing protein structure generative models and demonstrate that it accurately recovers complicated multi-modal alternate protein backbone conformations observed in certain single crystal measurements. 
\end{abstract}

\section{Introduction}
Proteins are dynamic molecules, transitioning between states during biological processes and otherwise thermodynamically sampling conformations within these states. Although models generated by X-ray crystallography typically depict a single conformation, this is actually an ensemble measurement.  Protein crystals are an enormous array of molecules and the electron density reconstructed from the diffraction captures variability between the atom locations in this array. As flexibility in the protein chain increases, the electron density becomes increasingly spread out.  Since it is difficult to recognise and model the specific conformations giving rise to the average density, variability around the best-fit model is typically reported only indirectly, in the form of the B factor. However, where detectable, crystallographers model atoms in multiple alternate locations (commonly termed, altlocs). Alternately located segments of the protein backbone have remained under-recognised, since most visualisation platforms (e.g., pymol and chimeraX) and programs using structural models as inputs (e.g., gromacs)  ignore altlocs altogether or resolve them with simple heuristics \cite{gutermuth2023modeling}. A recent work \cite{rosenberg2024seeingdouble} created a comprehensive catalogue of altlocs extracted from PDB structures, suggesting that this dataset should find use in efforts towards predicting multiple structures from a single sequence.  Interestingly, the authors showed that for a set of well-separated and stable altlocs, even if structural ensemble predictors recognise the region as being flexible, they fail to capture the experimentally determined conformations or even the bimodality of the distribution of backbone conformations. 

Representing proteins as ensembles of model structures, better capturing their dynamic nature, is a key goal in structural biology \cite{Wankowicz2023.06.28.546963} and protein structure prediction. We suggest an generative approach that directly produces samples that are faithful to the experimental data.  We showcase the proposed approach on electron densities resolved from X-ray crystallography, focusing on altloc regions having at least two clearly distinct conformations.

\paragraph{Contributions.} We formulate the problem of reconstructing a protein structure ensemble that obeys given experimental measurements as an inverse problem, in which the forward model describes the formation of the electron density given an ensemble. We use a pre-trained diffusion model to embody the prior joint probability of the atom coordinates given the amino acid sequence, and guide it to sample from the conditional probability given the observed density. We develop a differentiable forward model and a non-i.i.d. score guidance method that uses the entire ensemble sampled from the model in contrast to the standard score guidance that uses individual samples. 
%
%
We demonstrate that the proposed methodology faithfully recovers alternate conformations from crystallographic electron densities that are not reproduced correctly by unconditional sampling.

\section{Methods}\label{sec:methods}
\paragraph{Notation.} We denote by $\mathbf{a}$ the amino-acid sequence, $\mathbf{x}$ the backbone coordinates of a protein structure, and let $\boldsymbol{\chi}$ be the sidechain dihedral angles from which the side chain atom coordinates can be straightforwardly computed and denoted by $\{\mathbf{y}_i(\mathbf{x}, \boldsymbol{\chi}, \mathbf{a})\}$. Note that the dependence on the sequence $\mathbf{a}$ is implicit as the amino acid identities are required to calculate the atom locations from $\mathbf{x}$ and $\boldsymbol{\chi}$. We denote by $p(\mathbf{x})$ the distribution of the random variable $\mathbf{x}$, and by $F_\mathrm{o}: \mathbb{R}^3 \to \mathbb{R}$ and $F_\mathrm{c}: \mathbb{R}^3 \to \mathbb{R}$ the observed and calculated electron densities, respectively. 

\begin{wrapfigure}{r}{0.7\textwidth}
\centering
\vspace{-1cm}
\includegraphics[width=0.7\textwidth, trim=8mm 20mm 8mm 20mm, clip]{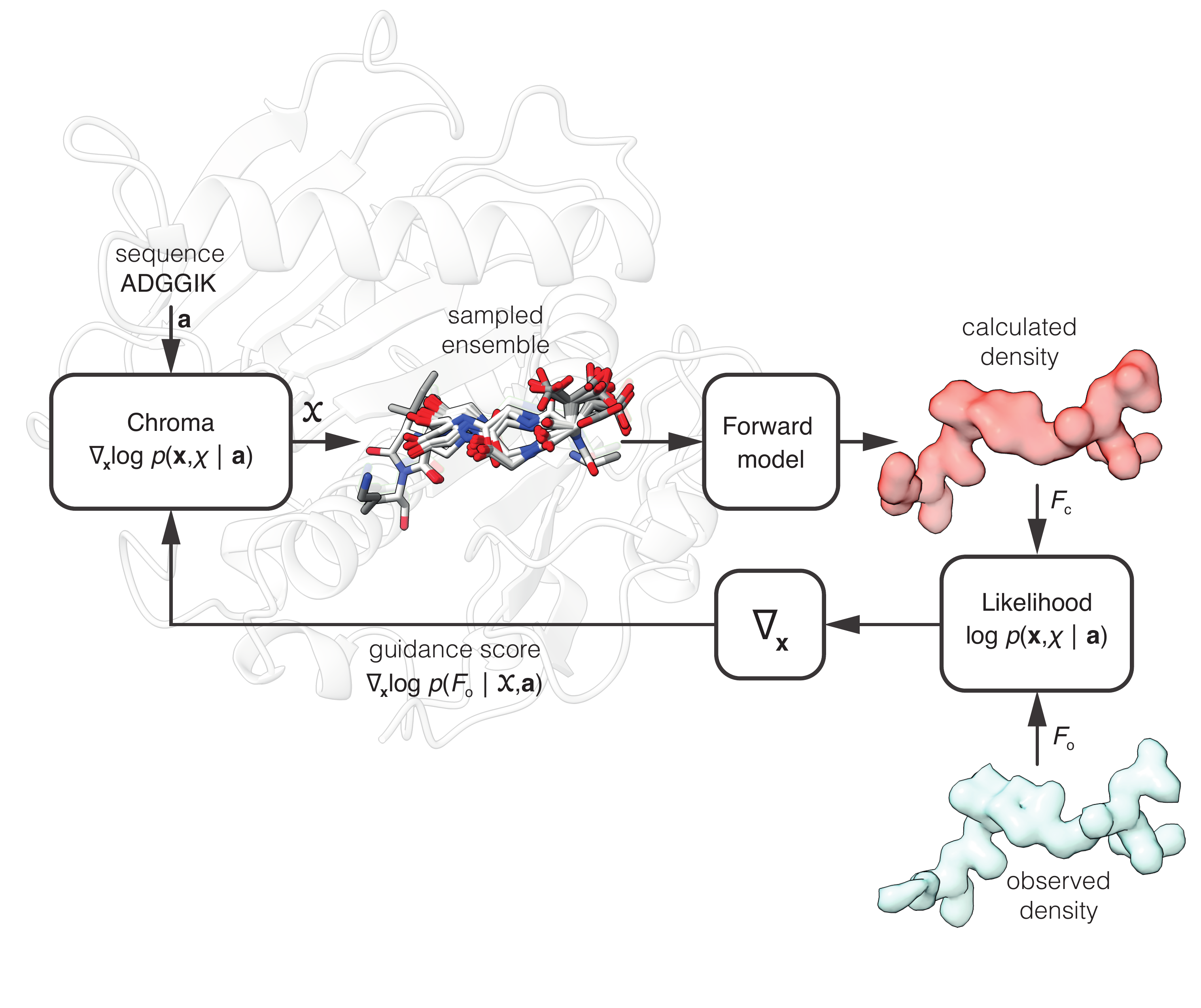}
\vspace{-0.6cm}
\caption{\textbf{The proposed density-guided protein ensemble generation method.} The diffusion model is used to sample a non-i.i.d. ensemble from which the likelihood of the observed electron density is calculated and used to guide the sampling. }
\vspace{-0.2cm}
\label{fig:our_guidance_approach}
\end{wrapfigure}

\paragraph{Problem setting.} Given an experimentally observed electron density of a protein structure $F_\mathrm{o}$ and the amino-acid sequence $\mathbf{a}$, our goal is to recover the posterior distribution $p(\mathbf{x}, \boldsymbol{\chi} | \mathbf{a}, F_\mathrm{o})$ of protein structure that explains $F_\mathrm{o}$ by sampling a \emph{non-i.i.d.} ensemble $\mathcal{X} = \{(\mathbf{x}^m, \boldsymbol{\chi}^m)\}$ from it. 
Following the standard Bayes' theorem, the posterior of the ensemble can be factorized as 
$p(\mathcal{X} | \mathbf{a}, F_\mathrm{o}) \propto p(F_\mathrm{o} | \mathcal{X}, \mathbf{a}) \cdot p(\mathcal{X} | \mathbf{a}) = 
p(F_\mathrm{o} | \mathcal{X}, \mathbf{a}) \cdot \prod_k p(\mathbf{x}^m, \boldsymbol{\chi}^m | \mathbf{a})$, where $p(F_\mathrm{o} | \mathcal{X}, \mathbf{a})$ is the \textit{likelihood} of observing $F_\mathrm{o}$ given a structure ensemble $\mathcal{X}$ and $p(\mathbf{x}, \boldsymbol{\chi} | \mathbf{a})$ is the \textit{prior} distribution of the protein conformations given the sequence. It is crucial to observe that the samples in the ensemble are not independent. While the prior can be split into independent terms $p(\mathbf{x}^m, \boldsymbol{\chi}^m | \mathbf{a})$, the likelihood generally depends on all the samples at once and is, therefore, inseparable. Our approach is summarized in Figure \ref{fig:our_guidance_approach}.

\paragraph{Protein structure prior.} We employ \chroma\,~\cite{Chroma2023} to model the distribution of feasible protein structures. While other protein structure generative models exist~\cite{bose2024se3stochastic, jing2024alphafold}, we chose \chroma\, because it is a score-based generative models, and it effectively models $p(\mathbf{x}, \boldsymbol{\chi}, \mathbf{a})$. We also experimented with distributional graphormer (DiG)\cite{zheng2024predicting}, however, we found that the score modeled by DiG is unstable, and it models only the backbone atoms, and does not model backbone oxygens and sidechains, which contribute to majority of the electron density.

\chroma\, models the all-atom representation of the protein structure by the following factorized form, $p(\mathbf{x}, \boldsymbol{\chi}, \mathbf{a}) = p(\mathbf{x}) p(\mathbf{a}, \boldsymbol{\chi} | \mathbf{x})$, where the distribution of backbone coordinates, $p(\mathbf{x})$, is modeled via a score-based diffusion model, and $p(\mathbf{a}, \boldsymbol{\chi} | \mathbf{x})$ is modeled as an autoregressive model. \chroma's forward diffusion process is modeled as a variance-preserving SDE \cite{song2019generative}, whose backward SDE is given by,
\begin{equation}\label{eq:uncond_sampling}
d\mathbf{x} = \left[-\frac{1}{2} \mathbf{x} - \mathbf{P}\mathbf{P}^\top \nabla_{\mathbf{x}} \log p_t(\mathbf{x}) \right] \beta_t d{t} + \sqrt{\beta_t} \mathbf{P} \mathbf{n},
\end{equation}
where $\mathbf{P}$ is a fixed preconditioning matrix, $\beta_t$ is the noise schedule, and $\mathbf{n}$ is isotropic Gaussian noise. The score function $\nabla_{\mathbf{x}}\log p_t (\mathbf{x})$ is modeled as an equivariant graph attention network. Sampling occurs by first drawing backbone coordinates by integrating Eq. \ref{eq:uncond_sampling} from $[T, 0]$. Modeling $\mathbf{x}_{\text{all}}$ given $\mathbf{x}$ is presented in the sequel. Because \chroma\ models the backbone coordinates $\mathbf{x}$ independently to the sequence, conditioning $\mathbf{x}_{\text{all}}$ on the sequence $\mathbf{a}$ can be done trivially by substitution. We discuss the details of sidechain packing in Section \ref{subsec:side_chain_packing}  in the Appendix.



\paragraph{Modeling the likelihood.} Assuming $p(F_\mathrm{o} | \mathcal{X}, \mathbf{a})$ is Gaussian, the log-likelihood of $F_\mathrm{o}$ is given by,
\begin{equation}
\log p(F_\mathrm{o} | \mathcal{X}, \mathbf{a}) = -\left\|F_\mathrm{o} - \frac{1}{|\mathcal{X}|} \sum_{m} F_\mathrm{c}\left(\mathbf{x}^m, \boldsymbol{\chi}^m, \mathbf{a}\right)\right \|_2^2,
\label{eq:guidance_term}
\end{equation}
where $\|\cdot\|_2$ is the $L_2$ norm in the space of electron densities, and the individual $F_\mathrm{c}$ terms are given by the kernel density estimates,
\begin{equation}
F_\mathrm{c}\left(\boldsymbol{\xi}\right) = \sum_{k=1}^{N_s} \sum_{i=1}^A \sum_{j=1}^5 a_{ij} \cdot \left(\frac{4\pi}{b_{ij} + B}\right)^{\frac{3}{2}} \cdot \exp\left(-\frac{4\pi^2}{b_{ij} + B} \cdot \|\mathbf{R}_k \mathbf{y}_i(\mathbf{x}^m, \boldsymbol{\chi}^m, \mathbf{a}) + \mathbf{t}_k - \boldsymbol{\xi}\|_2^2\right),
\end{equation}
where $N_s$ is the number of symmetry operations, $A$ is the number of atoms in the asymmetric unit, $\mathbf{R}_k$ is the rotation matrix of the $k$-th symmetry operation, $\mathbf{t}_k$ is the translation vector of the $k$-th symmetry operation, $\mathbf{y}_i$ is the location of the $i$-th atom, $a_{ij}$ and $b_{ij}$ are tabulated form factors \cite{prince2004international}, $B$ is the B-factor, and $\boldsymbol{\xi}$ is the point in space at which density is calculated. In standard molecular replacement and structure refinement pipelines, the B-factor is used to model the experimental electron density as a mixture of Gaussians. In our approach, however, we essentially sample directly from the joint distribution of the atom locations which allows more complex density models. The B-factor is, therefore, more akin to the bandwidth parameter in kernel density estimation (KDE) techniques. We apply a uniform B-factor across all atoms that should be inversely proportional to the ensemble size. Ideally, the optimization of the exact value should be part of the sampling procedure that we intend to develop in the future; here, we adopted a somewhat naive approach by taking the average of the individual B-factors for each atom in the given protein structure.


\paragraph{Sampling from the posterior.} To sample a non-independent ensemble from the posterior, we define the joint diffusion variable $\mathbf{X} = (\mathbf{x}^1, \dots, \mathbf{x}^{|\mathcal{X}|})$ concatenating all backbone coordinates of the ensemble, and plug Eq. \ref{eq:guidance_term} into the backward SDE equation \cite{dhariwal2021diffusion, Weiss2023},
\begin{equation}\label{eq:guided_sampling}
d\mathbf{X} = \left[-\frac{1}{2} \mathbf{X} - \mathbf{P}\mathbf{P}^\top ( \nabla_{\mathbf{X}} \log p_t(\mathbf{X})
+ \eta \cdot \nabla_{\mathbf{X}} \log p(F_\mathrm{o} | \mathbf{X}, \boldsymbol{\rchi}(\mathbf{X}), \mathbf{a})
)
\right] \beta_t d{t} + \sqrt{\beta_t} \mathbf{P} \mathbf{n}.
\end{equation}
Note that the diffusion is only applied to the backbone coordinates, from which the sidechain dihedrals  $\boldsymbol{\rchi}(\mathbf{X}) = (\boldsymbol{\chi}^1, \dots, \boldsymbol{\chi}^{|\mathcal{X}|})$ are calculated using the differentiable sidechain packer. While the unconditional (prior) score term, $\nabla_{\mathbf{X}} \log p_t(\mathbf{X})$, is separable (i.e., block diagonal), the guidance term is inseparable as mentioned above. The hyperparameter $\eta$ is used to scale the guidance score directing the diffusion model to generate samples that are better aligned with the observed density, thereby increasing the likelihood of Eq.~\ref{eq:guidance_term}.


\paragraph{Filtering samples using matching pursuit.} We observed that when sampling large ensembles using non-i.i.d. guidance, certain samples in the ensemble tend to overfit to noise in the density map, reducing the overall ensemble quality. To mitigate this, we adopt a matching pursuit-based approach \cite{mallat1993matching} to filter out low-quality samples and greedily select a subset of the ensemble, $\mathcal{X}_\mathcal{I} = \{ (\mathbf{x}^m, \boldsymbol{\chi}^m) : m \in \mathcal{I} \}$, that best aligns with $F_\mathrm{o}$. Starting with $\mathcal{I}=\emptyset$, each iteration seeks to maximize $\log p(F_\mathrm{o} | \mathcal{X}_{\mathcal{I} \cup \{ m \} }, \mathbf{a})$ over all $m \notin \mathcal{I}$. The optimal element $m$ is then added to the support set $\mathcal{I}$ and the process is repeated until the likelihood no longer increases. The procedure is summarized as Algorithm ~\ref{alg:filtering_pseudocode} in Appendix \ref{sec:appendix}.

\section{Results}\label{sec:results}
\paragraph{Experimental setup.} 
To evaluate the efficacy of our methods, we primarily aim to assess the alignment between the density computed from sampled conformers, $F_\mathrm{c}$, and the experimentally observed densities, $F_\mathrm{o}$. Additionally, we seek to investigate the ability of our approach to generate accurate conformers in the structural space. We, therefore, focus on regions where the density exhibits multi-modal behavior and aim to recover alternate locations (altlocs) consistent with those modeled in the original PDB structures.
To facilitate these investigations, we selected crystallographic protein structures from the Protein Data Bank (PDB) \cite{10.1093/nar/gkaa1038} that exhibit discernible separation between backbone and sidechain atoms at the altloc residues using the analysis in \cite{rosenberg2024seeingdouble}. Note that all these examples present highly intricate multi-modal backbone distributions that are poorly predicted by existing ensemble sampling techniques.  
We further restricted our attention to high resolution structures (2.5 \AA\ or better) to ensure high-quality electron density maps. 
Since the electron density maps available in the PDB are mean-centered and lack an absolute scale, we converted them to physical units ($e^{-} / \text{\AA}^{3}$) following the method described in \cite{doi:10.1073/pnas.1302823110}.
A comprehensive list of proteins used in our experiments is reported in Table \ref{tab:altloc_table}.

For all experiments, we will use the publicly available versions of \chroma's diffusion-based backbone sampler and \chroma's GNN-based sidechain packer for $\chi$-angle prediction as our prior \cite{Chroma2023}. Since the backbone sampler is not sequence-conditioned, applying density guidance across the entire protein would exacerbate the ill-posed nature of the problem. To mitigate this, we employ \chroma's \texttt{SubstructureConditioner} to constrain atoms outside the target residues to their ground truth locations.
All visualizations were rendered using ChimeraX \cite{pettersen2021ucsf}.

\begin{figure}
    \centering
    \includegraphics[width=.78\textwidth]{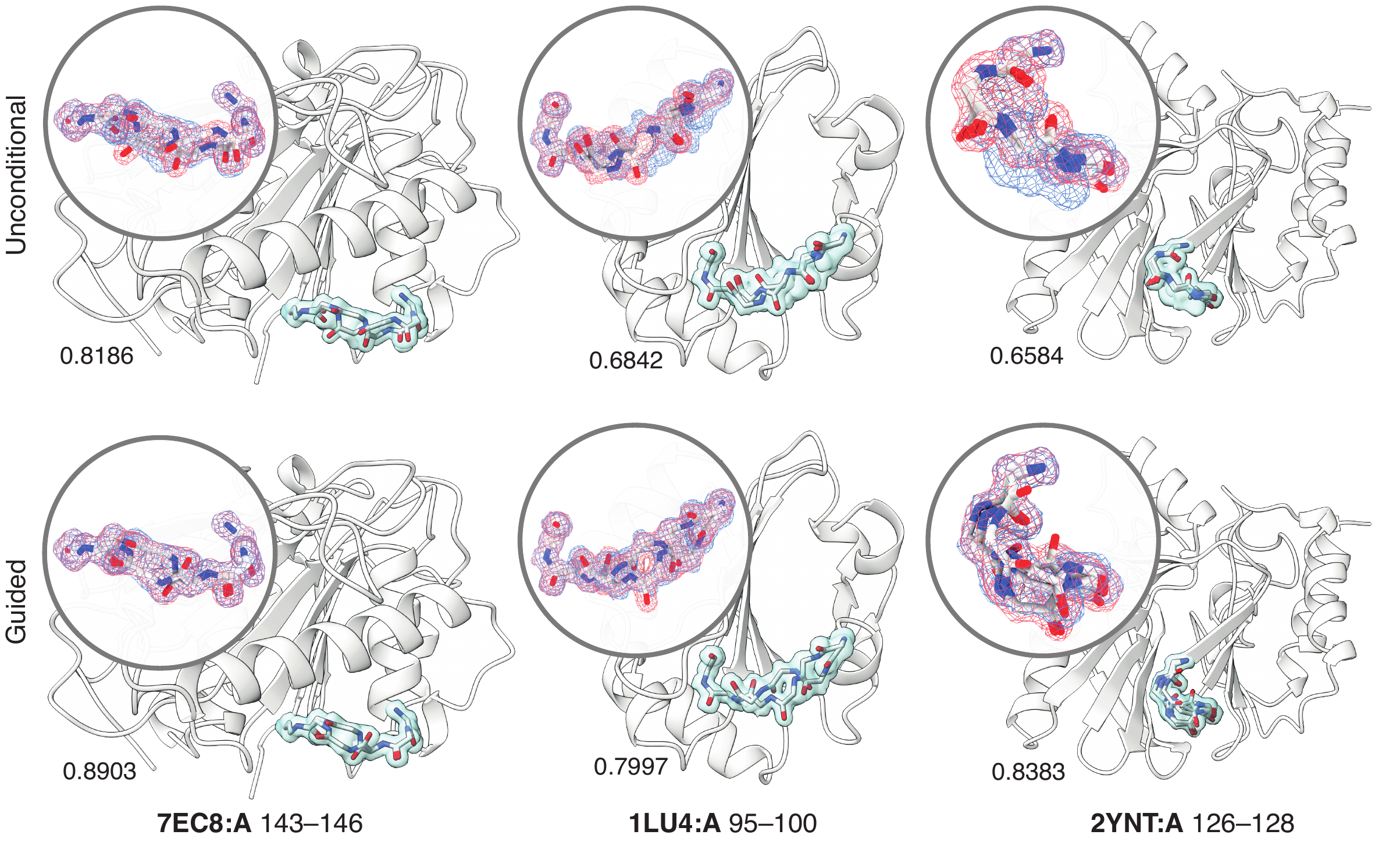}
    \caption{\textbf{Density-guided \texttt{Chroma} fits the density better than unguided \texttt{Chroma}, and recovers two known alternative locations accurately.} Full protein structure is displayed as white cartoons. The sampled ensembles in the region of interest are depicted as sticks and overlaid on the experimental density $1 \sigma$-isosurface. The inserts show the agreement of $F_\mathrm{c}$ (red) to the observed density $F_\mathrm{o}$ (blue) visualized as $1 \sigma$-isomeshes. Cosine similarities $F_\mathrm{c}$ and $F_\mathrm{o}$ are reported below each panel. Density guidance produces consistently better density alignment and correctly captures the multi-modal nature of the observed density. \vspace{-1mm}
    }
    \label{fig:qualitative_figure}
\end{figure}

\paragraph{Evaluation criteria.}
The results were evaluated using a batch sampling approach. For each batch, we identified a subset of samples that provided the optimal match to the observed electron density. The similarity between the observed electron density and the mean calculated density of the selected subset was quantified using cosine similarity $\langle F_\mathrm{o}, F_\mathrm{c} \rangle / (
\| F_\mathrm{o} \| \cdot \| F_\mathrm{c} \|)$ in the space of electron densities. 
Furthermore, we computed an alternate location (altloc) proximity score for each sample in the batch. The score was calculated as $\texttt{bimodality\_score}(\mathbf{x}) = \left(1 - \min( r (\mathbf{x}) , 1/r (\mathbf{x}) 
 )\right) \cdot \text{sign}(r (\mathbf{x}) - 1/r (\mathbf{x})  )$, where
$r (\mathbf{x}) = \|\mathbf{x} - \mathbf{x}_\mathrm{A}\| / \|\mathbf{x} - \mathbf{x}_\mathrm{B}\|
$,
and $\mathbf{x}_\mathrm{A}$ and $\mathbf{x}_\mathrm{B}$ are the backbone atom locations of altlocs A and B, respectively.  
In this formulation, $\text{score}(\mathbf{x})$ approaches $-1$ as $\mathbf{x}$ converges to altloc A, and approaches $+1$ as $\mathbf{x}$ converges to altloc B. 

\paragraph{Density alignment.} To evaluate the proposed method's effectiveness in aligning with $F_\mathrm{o}$, we compare the samples generated using density-guided diffusion with those from an unconditional diffusion process. Using the procedure outlined in Algorithm \ref{alg:sample_pseudo_code} we sample a batch of $16$ density-guided conformers. Likewise, a batch of $16$ unconditional conformers are sampled using Eq. \ref{eq:uncond_sampling}.
Both batches are filtered (Algorithm \ref{alg:filtering_pseudocode}) to retain at most $5$ samples that best fit the observed $F_\mathrm{o}$ map. From these best-performing conformers, we construct the $F_\mathrm{c}$ map and compute the correlation with $F_\mathrm{o}$ using cosine similarity. This evaluation is repeated across all proteins listed in Table $\ref{tab:altloc_table}$.
The quantitative results comparing unconditional sampling and density-guided sampling are presented in Figure $\ref{fig:cosine_similarities}$. Due to variations in resolution, the results are not comparable across proteins. However, we note that while \chroma\ provides a strong prior for protein structures, it struggles to accurately capture experimental density. This reinforces the finding that other protein ensemble sampling programs cannot correctly reproduce the observed densities in such intricately flexible regions \cite{rosenberg2024seeingdouble}.
Additionally, Figure \ref{fig:qualitative_figure} visually showcases, on three structures, how guided sampling improves the alignment between $F_\mathrm{c}$ and $F_{\mathrm{o}}$.

\begin{figure}[t] 
    \centering
    \vspace{-1mm}
    \includegraphics[width=.75\textwidth]{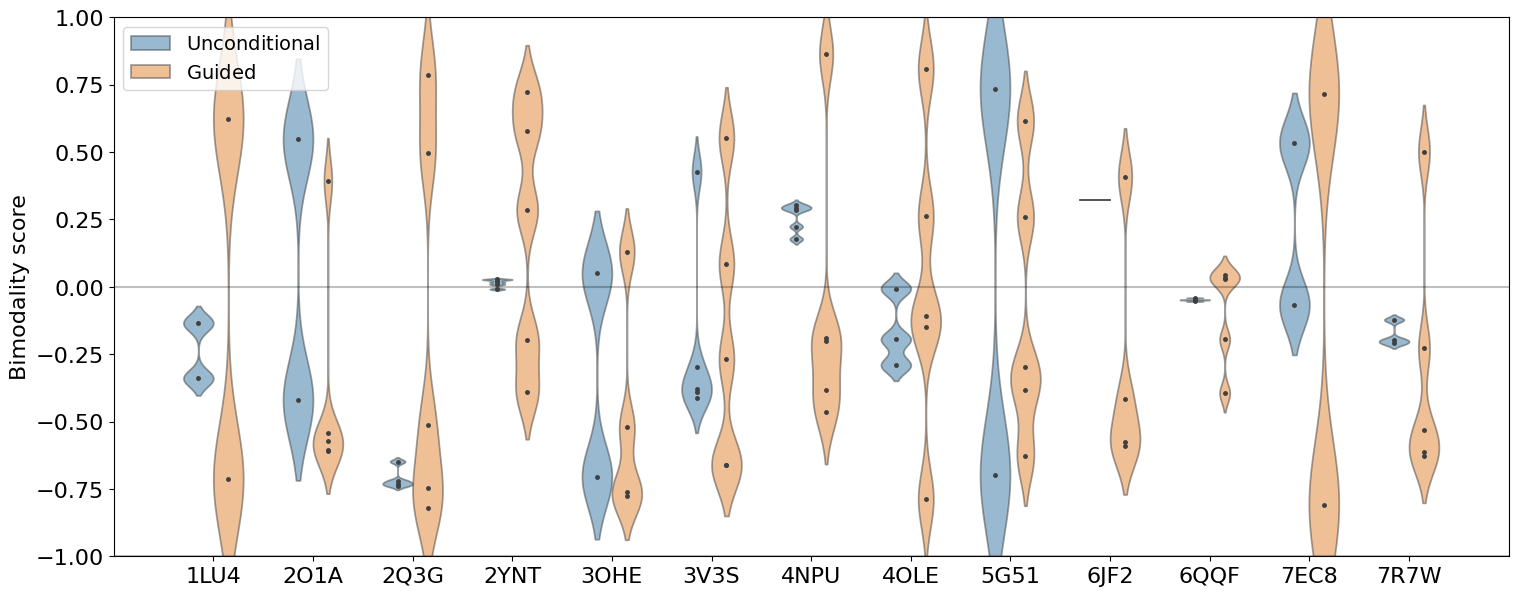} 
    \caption{\textbf{Density-guided \chroma\ accurately captures the bimodal distribution of the backbone conformations, while unconditional sampling consistently fails to represent it.} Negative scores represent proximity to the modeled altloc A, while positive scores correspond to altloc B. \vspace{-1mm}}
    \label{fig:altlocs_distances}
\end{figure}

\paragraph{Bimodality and structural alignment.} We evaluate the impact of aligning with the experimental density on the structural space. Specifically, we focus on regions of proteins with alternate conformations (altlocs) where the density exhibits a bimodal distribution. By conditioning our diffusion process on $F_\mathrm{o}$, we aim to capture this bimodality and generate samples consistent with the originally modeled altlocs in the PDB. From the aforementioned experiment, we utilize the filtered samples from both density-guided and unconditional ensembles. Each sample is assigned proximity score to one of the altloc (A or B). 
The resulting score distributions are visualized in Figure \ref{fig:altlocs_distances}. Guided sampling consistently achieves bi- and multi-modal behavior with proximities to both modeled altlocs (positive and negative modes in the plot), while the unconditional counterpart often fails to correctly represent this behavior.
Additionally, our method performs well in regions with unimodal density and no alternative conformations, as detailed in Section \ref{sec:unimodal} and visualized in Figures~\ref{fig:unimodal_controls_plots} and \ref{fig:unimodal_density_comparison} in the Appendix.





\section{Conclusion}\label{sec:conclusion}
In this work, we presented a novel density-guided generative modeling approach for reconstructing protein structure ensembles from crystallographic electron density maps. By formulating the task as an inverse problem, we leveraged a pre-trained diffusion model as a flexible prior over protein backbone conformations, introducing a non-i.i.d. score guidance technique that optimizes the ensemble as a whole, rather than individual structures thereof. Our method outperforms unconditional sampling, particularly in multimodal regions, recovering altlocs consistent with PDB data and aligning more closely with the observed electron densities. Looking forward, our method opens several promising directions for future exploration, such as extending to larger proteins, enhancing the modeling of B-factors, improving sidechain packing accuracy, and extending to the forward models of other experimental modalities such as cryoEM/ET. We believe that our work also provides an important step toward more accurate and data-driven modeling of protein dynamics.


\clearpage
 \bibliographystyle{plain} \bibliography{neurips_2024}

\begin{thebibliography}{10}

\bibitem{bose2024se3stochastic}
Avishek~Joey Bose, Tara Akhound-Sadegh, Guillaume Huguet, Killian Fatras, Jarrid Rector-Brooks, Cheng-Hao Liu, Andrei~Cristian Nica, Maksym Korablyov, Michael Bronstein, and Alexander Tong.
\newblock Se(3)-stochastic flow matching for protein backbone generation.
\newblock In {\em The International Conference on Learning Representations (ICLR)}, 2024.

\bibitem{10.1093/nar/gkaa1038}
Stephen~K Burley, Charmi Bhikadiya, Chunxiao Bi, Sebastian Bittrich, Li~Chen, Gregg~V Crichlow, Cole~H Christie, Kenneth Dalenberg, Luigi Di~Costanzo, Jose~M Duarte, et~al.
\newblock Rcsb protein data bank: powerful new tools for exploring 3d structures of biological macromolecules for basic and applied research and education in fundamental biology, biomedicine, biotechnology, bioengineering and energy sciences.
\newblock {\em Nucleic acids research}, 49(D1):D437--D451, 2021.

\bibitem{dhariwal2021diffusion}
Prafulla Dhariwal and Alexander Nichol.
\newblock Diffusion models beat gans on image synthesis.
\newblock {\em Advances in neural information processing systems}, 34:8780--8794, 2021.

\bibitem{gutermuth2023modeling}
Torben Gutermuth, Jochen Sieg, Tim Stohn, and Matthias Rarey.
\newblock Modeling with alternate locations in x-ray protein structures.
\newblock {\em Journal of Chemical Information and Modeling}, 63(8):2573--2585, 2023.

\bibitem{Chroma2023}
John~B. Ingraham, Max Baranov, Zak Costello, Karl~W. Barber, Wujie Wang, Ahmed Ismail, Vincent Frappier, Dana~M. Lord, Christopher Ng-Thow-Hing, Erik~R. Van~Vlack, Shan Tie, Vincent Xue, Sarah~C. Cowles, Alan Leung, Jo\~{a}o~V. Rodrigues, Claudio~L. Morales-Perez, Alex~M. Ayoub, Robin Green, Katherine Puentes, Frank Oplinger, Nishant~V. Panwar, Fritz Obermeyer, Adam~R. Root, Andrew~L. Beam, Frank~J. Poelwijk, and Gevorg Grigoryan.
\newblock Illuminating protein space with a programmable generative model.
\newblock {\em Nature}, 2023.

\bibitem{jing2024alphafold}
Bowen Jing, Bonnie Berger, and Tommi Jaakkola.
\newblock Alphafold meets flow matching for generating protein ensembles.
\newblock In {\em Forty-first International Conference on Machine Learning}, 2024.

\bibitem{doi:10.1073/pnas.1302823110}
P.~Therese Lang, James~M. Holton, James~S. Fraser, and Tom Alber.
\newblock Protein structural ensembles are revealed by redefining x-ray electron density noise.
\newblock {\em Proceedings of the National Academy of Sciences}, 111(1):237--242, 2014.

\bibitem{mallat1993matching}
St{\'e}phane~G Mallat and Zhifeng Zhang.
\newblock Matching pursuits with time-frequency dictionaries.
\newblock {\em IEEE Transactions on signal processing}, 41(12):3397--3415, 1993.

\bibitem{pettersen2021ucsf}
Eric~F Pettersen, Thomas~D Goddard, Conrad~C Huang, Elaine~C Meng, Gregory~S Couch, Tristan~I Croll, John~H Morris, and Thomas~E Ferrin.
\newblock Ucsf chimerax: Structure visualization for researchers, educators, and developers.
\newblock {\em Protein science}, 30(1):70--82, 2021.

\bibitem{prince2004international}
Edward Prince.
\newblock {\em International Tables for Crystallography, Volume C: Mathematical, physical and chemical tables}.
\newblock Springer Science \& Business Media, 2004.

\bibitem{rosenberg2024seeingdouble}
Aviv~A. Rosenberg, Sanketh Vedula, Alex~M. Bronstein, and Ailie Marx.
\newblock Seeing double: Molecular dynamics simulations reveal the stability of certain alternate protein conformations in crystal structures.
\newblock {\em bioRxiv}, 2024.

\bibitem{shapovalov2011smoothed}
Maxim~V Shapovalov and Roland~L Dunbrack.
\newblock A smoothed backbone-dependent rotamer library for proteins derived from adaptive kernel density estimates and regressions.
\newblock {\em Structure}, 19(6):844--858, 2011.

\bibitem{song2019generative}
Yang Song and Stefano Ermon.
\newblock Generative modeling by estimating gradients of the data distribution.
\newblock {\em Advances in neural information processing systems}, 32, 2019.

\bibitem{van2008partially}
David~A Van~Dyk and Taeyoung Park.
\newblock Partially collapsed gibbs samplers: Theory and methods.
\newblock {\em Journal of the American Statistical Association}, 103(482):790--796, 2008.

\bibitem{Wankowicz2023.06.28.546963}
Stephanie~A. Wankowicz, Ashraya Ravikumar, Shivani Sharma, Blake~T. Riley, Akshay Raju, Daniel~W. Hogan, Henry van~den Bedem, Daniel~A. Keedy, and James~S. Fraser.
\newblock Uncovering protein ensembles: Automated multiconformer model building for x-ray crystallography and cryo-em.
\newblock {\em bioRxiv}, 2023.

\bibitem{Weiss2023}
Tomer Weiss, Eduardo Mayo~Yanes, Sabyasachi Chakraborty, Luca Cosmo, Alex~M. Bronstein, and Renana Gershoni-Poranne.
\newblock Guided diffusion for inverse molecular design.
\newblock {\em Nature Computational Science}, 3(10):873--882, Oct 2023.

\bibitem{zheng2024predicting}
Shuxin Zheng, Jiyan He, Chang Liu, Yu~Shi, Ziheng Lu, Weitao Feng, Fusong Ju, Jiaxi Wang, Jianwei Zhu, Yaosen Min, et~al.
\newblock Predicting equilibrium distributions for molecular systems with deep learning.
\newblock {\em Nature Machine Intelligence}, pages 1--10, 2024.

\end{thebibliography}
\clearpage
\appendix
\section{Appendix and Supplemental Material}\label{sec:appendix}

\renewcommand\thefigure{\thesection\arabic{figure}}
\renewcommand\thetable{\thesection\arabic{table}}
\renewcommand\thealgorithm{\thesection\arabic{algorithm}}
\setcounter{figure}{0}  
\setcounter{table}{0}  
\setcounter{algorithm}{0} 

\subsection{Side-chain packing}\label{subsec:side_chain_packing} As previously mentioned in Section \ref{sec:methods}, the diffusion model's prior is defined over the backbone coordinates of the protein structure. However, majority of the electron density is concentrated near the sidechain atoms. To account for this, we leverage \chroma's sidechain packer to sample the $\chi$ dihedral angles and impute the sidechain atoms using the sampled angles and the rotamer library \cite{shapovalov2011smoothed}. Since the density map is generated from a noise-free protein structure, we denoise the backbone atoms $\mathbf{x}$ at each iteration using the diffusion model before computing the $\chi$ angles. The denoised variable $\hat{\mathbf{x}}_0$ serves as an estimate of the noiseless backbones $\mathbf{x}$.

The $\chi$ sampler is an autoregressive sidechain decoder that models $l_{\theta} (\boldsymbol{\chi} | \mathbf{x}, \mathbf{a}) \approx \log p (\boldsymbol{\chi} | \mathbf{x})$, where $\theta$ denote the model parameters. The autoregressive decomposition of the likelihood is given by
\begin{equation*}
    l_{\theta} (\boldsymbol{\chi} | \mathbf{x}, \mathbf{a}) = \sum_{i} l_{\theta} (\chi_{i} | \chi_{i-1}, \chi_{i-2}, \dots \chi_{1}, \mathbf{a}, \mathbf{x}).
\end{equation*}
Once the $\chi$ angles are sampled, the combination with the backbone coordinates $\mathbf{x}$ fully describes the protein structure. 

Unfortunately, sampling $\chi$ angles for every amino acid at each timestep of the diffusion process introduces a significant computational bottleneck, which can severely impede the sampling procedure. To mitigate this, we restrict $\chi$ dihedral angle sampling to the amino acids being optimized, along with a window of size $W=5$ amino acids before and after the target region. Incorporating this window provides the model with greater structural context, thereby enhancing the accuracy of the angle predictions. For non-target residues, where the structure is anchored by the \texttt{SubstructureConditioner}, $\chi$ angles are computed using the ground truth coordinates from the PDB file.

Additionally, performing $\chi$ dihderal angle sampling and backpropagating through the $\chi$ sampling and the diffusion models at every iteration results in an excessively long and computationally expensive backpropagation graph. Given that atomic coordinates generally exhibit minimal change between iterations, this repeated sampling and backpropagation can become redundant. Hence, we implement a Gibbs-sampling-based optimization strategy \cite{van2008partially}.

Specifically, we sample the dihedral angles only once every $T_{\chi}$ iterations and then detach them from the backpropagation graph. During the intermediate iterations, we freeze the backbone coordinates $\mathbf{x}$ and optimize the angles $S_{\chi}$ times based on Eq. \ref{eq:guidance_term}. This approach stabilizes the backpropagtaion process while ensuring that the sidechains conform to the observed density $F_\mathrm{o}$.

The specifics of sidechain packing are presented in Algorithm \ref{alg:sample_pseudo_code}.

\subsection{Hyperparameters}\label{subsec:hyperparameters}
In our experiment, we employed a guidance scale of $9.0$ and clipped the gradient norms to a maximum of $32.5$. The $\chi$ diheral angels were resampled every 40 diffusion steps ($T_{\chi}$). For each diffusion step, 25 SGD optimization steps ($S_{\chi}$) were performed on the $\chi$ with a step size of $0.001$. We generated a batch of 16 samples with non-i.i.d guidance. Notably, all the results presented in this paper were obtained using these hyperparameters without any additional tuning, highlighting the simplicity and robustness of our density-guidance approach.

\subsection{Unimodal densities}\label{sec:unimodal}
\begin{figure}[t]
    \centering
    \includegraphics[width=1.\textwidth]{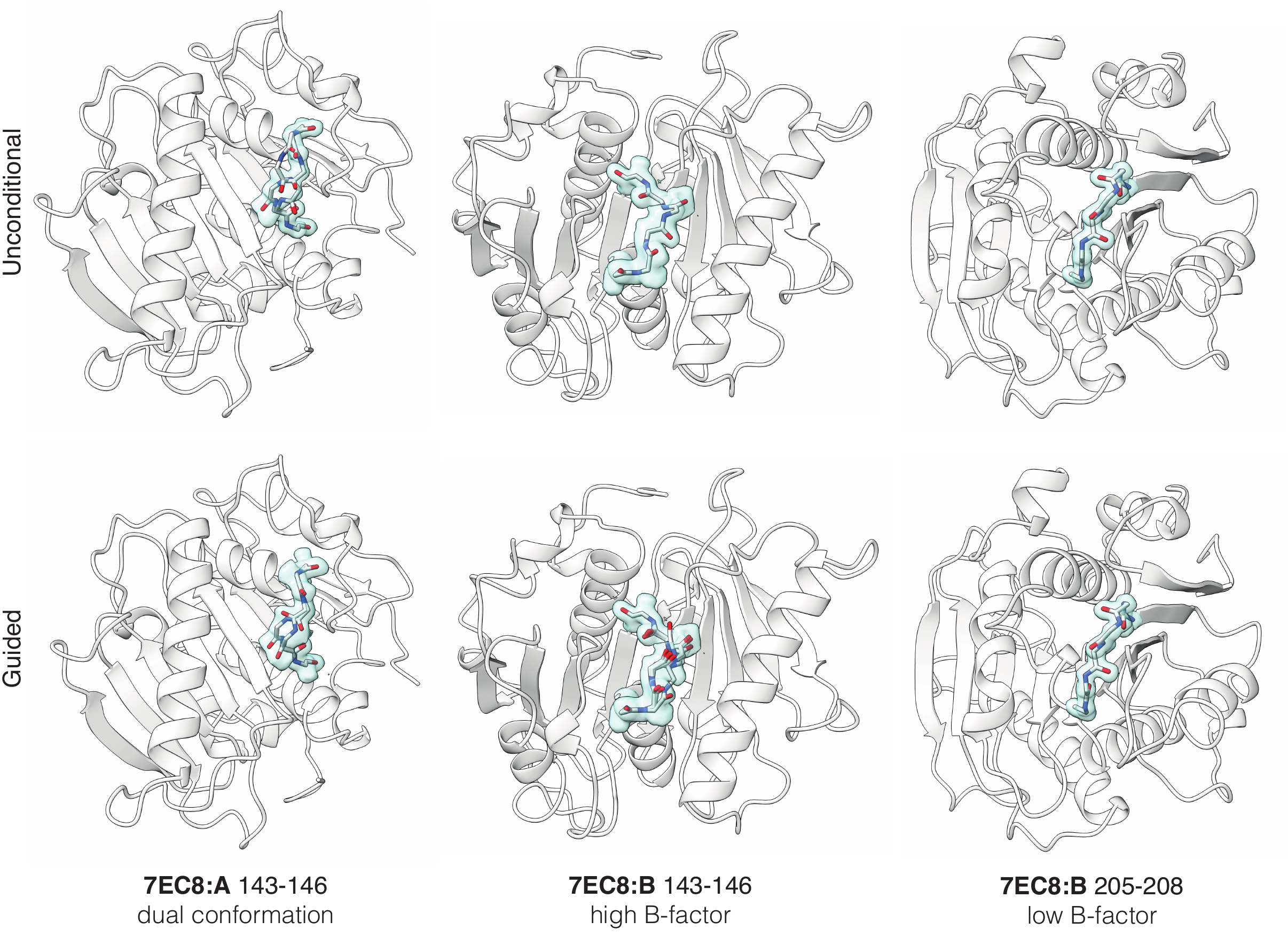}
    \caption{\textbf{Comparison of conditional and unconditional sampling in bimodally- and unimodally-distributed regions of protein \texttt{7EC8:A}}. The figure illustrates the differences between unconditional sampling (first row) and density-guided (second row) sampling methods in three regions of the protein (left-to-right): residues $143-146$ of chain A exhibiting an explicitly modeled dual conformation (two altlocs), the same position in chain B originally modeled as a high B-factors single conformation, and residues $205-208$ in chain B, originally modeled as a low B-factors single conformation. Our density-guided sampling consistently describes the flexible region in both chains as a bimodal distribution while producing a tightly distributed ensemble for the third low B-factor region.}
    \label{fig:unimodal_controls_plots}
\end{figure}

\begin{figure}
    \centering
    \includegraphics[width=0.7\textwidth]{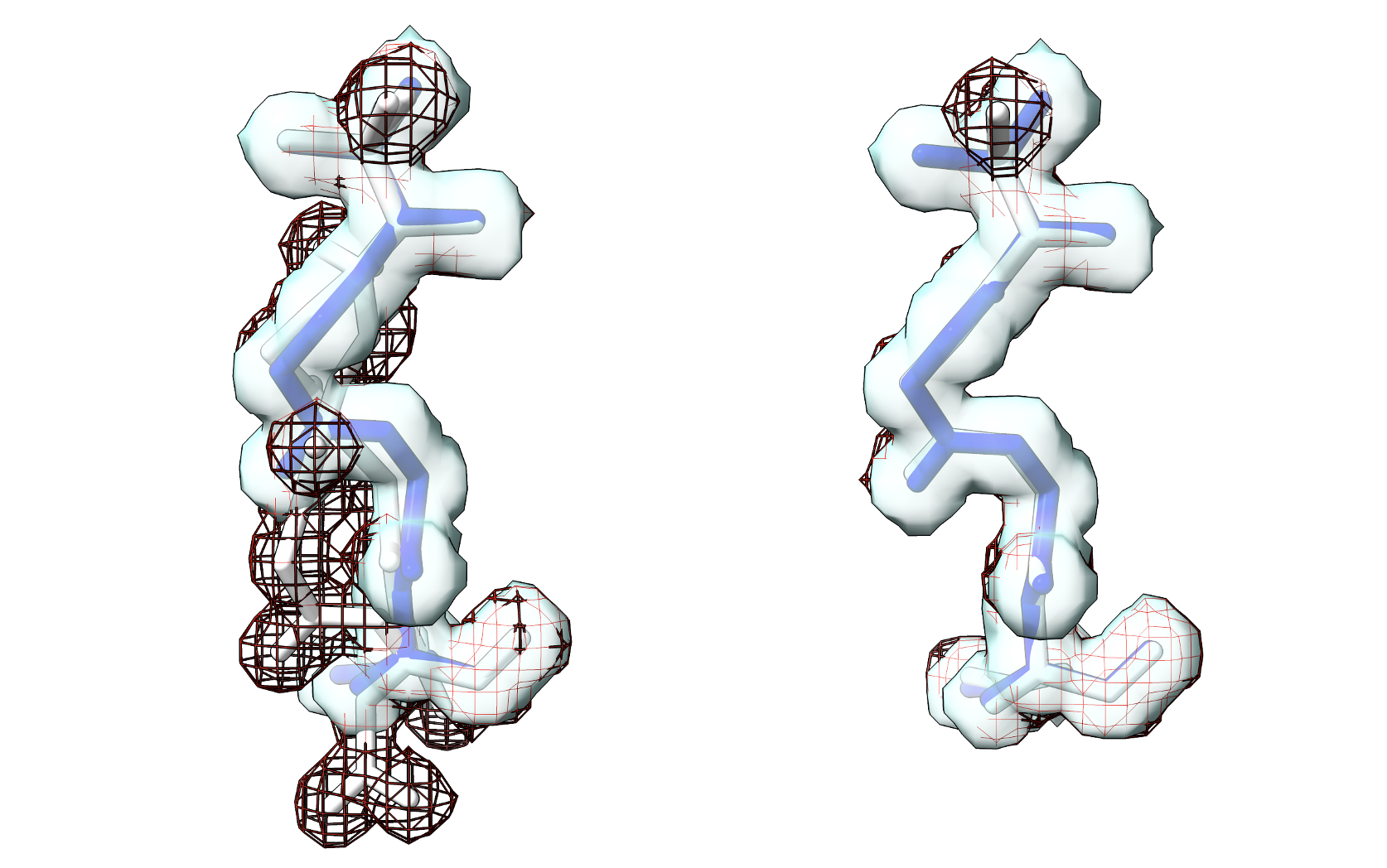}
    \caption{
    \textbf{Comparison of density fitting for residues 205-208 of protein \texttt{7EC8}}. The unguided approach (left) does not adequately align the calculated density with the observed density, whereas the guided approach (right) demonstrates significantly better alignment. The observed density is represented by the light blue surface, while the calculated density from the samples is shown as a dark red isomesh.
}
    \label{fig:unimodal_density_comparison}
\end{figure}

As a control experiment, we evaluated our method on protein 7EC8, specifically on chain B, which exhibits unimodal behavior. We sampled two regions: residues 205-208, which are characterized by a low B-factor, and residues 143-146, which display a higher B-factor. Although both regions are unimodal, the increased B-factor in residues 143-146 led to more spread-out and blurred density, resulting in slight variability in the samples. The cosine similarity results for both regions are shown in Table~\ref{tab:control_results}, demonstrating that our guided method consistently outperformed the unconditional approach. A visual comparison of the results is presented in Figures~\ref{fig:unimodal_controls_plots} and~\ref{fig:unimodal_density_comparison}.

\begin{table}[htbp]
\centering
\begin{tabular}{cccc}
\toprule
\textbf{Residue Range} & \textbf{Guided} & \textbf{Unconditional} \\ 
\midrule
205--208 & 0.9240 & 0.7886 \\
143--146 & 0.9316 & 0.7814 \\
\bottomrule\\
\end{tabular}
\caption{Cosine similarity results for two sampled regions in chain B of protein \texttt{7EC8}.}
\label{tab:control_results}
\end{table}

\begin{figure}[h] 
    \centering
    \includegraphics[width=1.\textwidth]{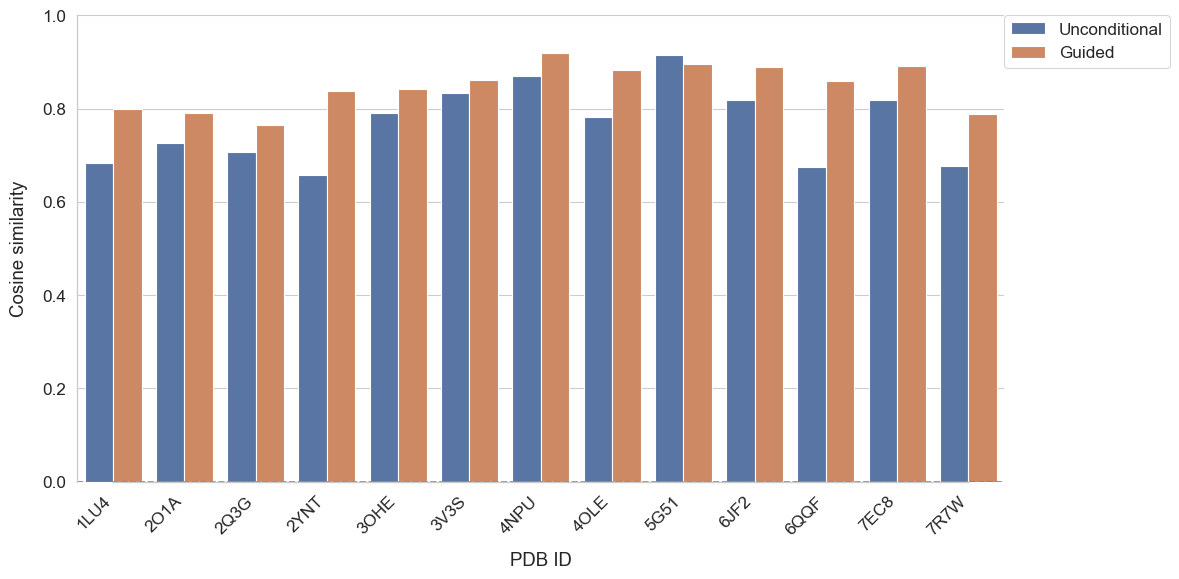} 
    \caption{\textbf{Comparison of cosine similarities between ensembles obtained using unconditional and density-guided sampling.}}
    \label{fig:cosine_similarities}
\end{figure}

\subsection{Ablation}
We conducted an ablation study on the batched (non-i.i.d) sampling procedure. As discussed in Section \ref{sec:methods}, our density-guidance approach leverages a batch of non-i.i.d samples, where we compute an expectation over $F_\mathrm{c}$ maps and compare it to $F_\mathrm{o}$ as depicted in Eq. \ref{eq:guidance_term}. In contrast, when we generate i.i.d samples by optimizing each sample individually without taking an expectation -- the performance deteriorates, as shown in Table \ref{tab:ablation_non_iid}.

\begin{table}[htbp]
\centering
\begin{tabular}{ccc}
\toprule
\textbf{PDB} & \textbf{Guided (batched)} & \textbf{Guided (i.i.d)} \\
\midrule
\texttt{2O1A} & \textbf{0.7903} & 0.7459 \\
\texttt{3OHE} & \textbf{0.8416} & 0.8369 \\
\texttt{7EC8} & \textbf{0.8903} & 0.8585 \\
\bottomrule
\vspace{2mm}
\end{tabular}
\caption{\textbf{Batched sampling of protein ensembles produces superior alignment to $F_{\mathrm{o}}$ compared to i.i.d sampling}.}
\label{tab:ablation_non_iid}
\end{table}

\clearpage
\subsection{Algorithms}\label{subsec:algorithms}

\begin{minipage}{1.0\textwidth}
\begin{algorithm}[H]
    \caption{Density-guided backbone sampling}\label{alg:sample_pseudo_code}
    \begin{algorithmic}[1]
        \Require Backbone denoising diffusion model $\epsilon_{\theta}$; $\chi$ sampling model $\chi_{\theta}$; $\chi$ resample time steps $T_{\chi}$; number of $\chi$ optimization steps $S_{\chi}$; $\chi$ optimization step size $\lambda_{\chi}$; target density $F_{\mathrm{o}}$; number of diffusion steps $N$; guidance scale $\lambda$; preconditioning matrix $\mathbf{P}$; diffusion schedule constants $\{\alpha_n\}_{n=1}^{N}$, $\{\beta_n\}_{n=1}^{N}$ as in \texttt{chroma}; amino acid sequence of the protein $\mathbf{a}$; $F_{\mathrm{c}}$ density forward model.
        
        \State $\mathbf{z} \sim \mathcal{N}(\mathbf{0}, \mathbf{I})$
        \State $\mathbf{x}_N = \mathbf{P} \mathbf{z}$
        \For{$n = N$ \textbf{to} $1$}
            \State $\mathbf{n} \sim \mathcal{N}(\mathbf{0}, \mathbf{I})$
            \State $\hat{\epsilon} = \epsilon_{\theta}(\mathbf{x}_n, \mathbf{n})$
            \State $\hat{\mu} = \frac{\alpha_n + 1}{2(1 - \alpha_n)} \mathbf{x}_n - \frac{\sqrt{\alpha_n}}{1 - \alpha_n} \hat{\epsilon}$
            \State $\hat{\mathbf{x}}_0 = \dfrac{\mathbf{x}_n - \sqrt{1 - \alpha_n} \hat{\epsilon}}{\sqrt{\alpha_n}}$
            \If{$n \in T_{\chi}$}
                \State $\hat{\chi}_n = \chi_{\theta}(\hat{\mathbf{x}}_0)$
            \Else \Comment{Optimize the side-chain angles}
                \State Initialize $\tilde{\chi}_{n, 0} = \hat{\chi}_{n + 1}$
                \For{$i = 0$ \textbf{to} $S_{\chi} - 1$}
                    \State $\tilde{\chi}_{t, i + 1} = \tilde{\chi}_{t, i} - \lambda_{\chi} \nabla_{\chi} \left\| F_{\mathrm{c}}(\texttt{all\_atom}(\hat{\mathbf{x}}_0, \tilde{\chi}_{t, i}, \mathbf{a})) - F_{\mathrm{o}} \right\|_2^2$
                \EndFor
                \State $\hat{\chi}_n = \tilde{\chi}_{n, S_{\chi}}$
            \EndIf
            \State $\mathbf{x}_{n - 1} = \hat{\mu} + \lambda \nabla_\mathbf{x} \left[ \left\| F_{\mathrm{c}}( \texttt{all\_atom}(\hat{\mathbf{x}}_0, \hat{\chi}_n, \mathbf{a})) - F_{\mathrm{o}} \right\|_2^2\right] + \sqrt{\beta_n} \mathbf{P} \mathbf{n}$ \Comment{Eq. \ref{eq:guided_sampling}}
        \EndFor
        \State \Return $\mathbf{x}_0$
    \end{algorithmic}
\end{algorithm}
\end{minipage}

\begin{algorithm}
    \caption{Selecting samples using matching pursuit \cite{mallat1993matching}}
    \label{alg:filtering_pseudocode}
    \begin{algorithmic}[1]
    \Require
        $\mathbf{D} = \{d_1, \ldots, d_n\}$: Sample densities ($F_\mathrm{c}$) from forward model, $d_t$: Target density from $F_\mathrm{o}$ density map, $\text{corr}(\cdot, \cdot)$: Function to compute correlation metric, $m_{\text{max}}$: Maximum allowed samples to select, $\mathbf{a}$: Amino Acid sequence of the protein
    
        
        \State $\mathcal{I} = \emptyset$ 
        \State $\mathcal{P} = \{0, 1, 2, \dots, |D| - 1\}$
        \State $s_{\text{current}} = 0$
        \While{$|\mathcal{I}| < m_{\text{max}}$}
            \State $\mathbf{c} = \{\text{corr}(d_{\mathcal{I} \cup \{i\}}, d_t) ~|~ d_{i} \in \mathbf{D}, i \notin \mathcal{P} \}$
            \State $i_{\text{best}}, s_{\text{max}} = \underset{i}{\text{argmax }} \mathbf{c}, \max \mathbf{c}$ \Comment{Maximize $\log p(F_\mathrm{o} | \mathcal{X}_{\mathcal{I} \cup \{ i_{\text{best}} \} }, \mathbf{a})$}
            \State $\mathcal{I} = \mathcal{I} \cup \{i_{\text{best}}\}$ \Comment{Add best sample}
            \State $\mathcal{P} = \mathcal{P} \setminus \{i_{\text{best}}\}$ \Comment{Update candidates}
            \If{$s_{\text{max}} < s_{\text{current}}$}
            \State break
            \EndIf
            \State $s_{\text{current}} = s_{\text{max}}$
        \EndWhile
        \State \Return $\mathcal{I}$


    \end{algorithmic}
\end{algorithm}
\clearpage
\subsection{Dataset}
\begin{table}[h!]
\centering
\resizebox{\textwidth}{!}{ 
\begin{tabular}{>{\centering\arraybackslash}p{2cm} 
                >{\centering\arraybackslash}p{3.5cm} 
                >{\centering\arraybackslash}p{5cm} 
                >{\centering\arraybackslash}p{2.5cm} 
                >{\centering\arraybackslash}p{2.5cm} 
                >{\centering\arraybackslash}p{2.5cm}
                >{\centering\arraybackslash}p{2.5cm}}
                
\toprule
\textbf{PDB ID} & \textbf{Protein function} & \textbf{Source organism} & \textbf{Expr. system} & \textbf{Altloc site} & \textbf{Altloc sequence}  & \textbf{Resolution (\AA)}\\
\midrule
\texttt{1LU4:A}  & Oxidoreductase  & Mycobacterium tuberculosis  & E. coli   & 95 -- 100    & \texttt{YNVPWW} & 1.12 \\
\addlinespace
\texttt{2O1A:A}  & Surface active protein  & Staphylococcus aureus  & E. coli   & 50 -- 53    & \texttt{KQNN} & 1.60\\
\addlinespace
\texttt{2Q3G:A}  & Structural Genomics  & Homo sapiens  & E. coli   & 24 -- 27    & \texttt{FNVP} & 1.11\\
\addlinespace
\texttt{2YNT:A}  & Hydrolase               & Pseudomonas aeruginosa & E. coli   & 126 -- 128  & \texttt{GNG} & 1.60\\
\addlinespace
\texttt{3OHE:A}  & Hydrolase               & Marinobacter nauticus VT8  & E. coli & 97 -- 104   & \texttt{YQGDPAWP} & 1.20\\
\addlinespace
\texttt{3V3S:B}  & Hydrolase               & Pseudomonas aeruginosa   & E. coli  & 227 -- 232  & \texttt{KAQERD} & 1.90\\
\addlinespace
\texttt{4NPU:B}  & Hydrolase               & Human immunodeficiency virus 1  & E. coli  & 33 -- 36  & \texttt{FEEI} & 1.50\\
\addlinespace
\texttt{4OLE:B}  & Unknown               & Homo sapiens  & E. coli  & 59 -- 69  & \texttt{ASTEKKDVLVP} & 2.52\\
\addlinespace

\texttt{5G51:A}  & Viral Protein               & Deformed wing virus  & E. coli  & 31 -- 36  & \texttt{GSASDQ} & 1.45\\
\addlinespace
\texttt{6JF2:A}  & Structural protein      & \makecell{Salmonella enterica subsp.\\ enterica serovar Typhimurium \\ str. LT2}  & E. coli  & 50 -- 54  & \texttt{VTAGG}    & 2.00\\
\addlinespace
\texttt{6QQF:A}  & Hydrolase      & Gallus gallus  & Gallus gallus  & 68 -- 75  & \texttt{RTPGSRNL}     & 1.95\\
\addlinespace
\texttt{7EC8:A}  & Hydrolase               & Uncultured bacterium     & E. coli  & 187 -- 190  & \texttt{DGGI}     & 1.35\\
\addlinespace
\texttt{7R7W:B}  & Immune system           & Homo sapiens             & E. coli  & 46 -- 50    & \texttt{IEKVE}   & 1.17\\
\bottomrule
\vspace{2mm}
\end{tabular}
}
\caption{ \textbf{Protein structures used for the evaluation in our experiments.} Note that the residue indices of the altloc sites are given with respect to the first residue in the structure (numbered as $1$) and may differ from the author-assigned indices in the original structures. }
\label{tab:altloc_table}
\end{table}
\end{document}